\documentclass[prl,twocolumn,showpacs,superscriptaddress,
preprintnumbers,amsmath,amssymb,tightenlines]{revtex4}
\usepackage{graphicx}
\newcommand{\beq}{\begin{equation}}
\newcommand{\eeq}{\end{equation}}
\newcommand{\bea}{\begin{eqnarray}}
\newcommand{\eea}{\end{eqnarray}}
\newcommand{\ba}{\begin{array}}
\newcommand{\ea}{\end{array}}
\newcommand{\bc}{\begin{center}}
\newcommand{\ec}{\end{center}}

\newcommand{\bml}{\begin{mathletters}}
\newcommand{\eml}{\end{mathletters}}
\newcommand{\commentout}[1]{{}}

\newcommand{\half}{\hbox{$1\over2$}}

\newcommand{\comment}[1]{{}}

\begin{document}
\title{Split-step Fourier methods for the Gross-Pitaevskii
equation}
\author{Juha Javanainen}
\affiliation{Department of Physics, University of Connecticut,
Storrs, Connecticut 06269-3046}
\email{jj@phys.uconn.edu}
\author{Janne Ruostekoski}
\affiliation{Department of Physics, Astronomy and Mathematics,
University of Hertfordshire, Hatfield, Herts, AL10 9AB, UK}
\affiliation{Institute for Theoretical Atomic and Molecular Physics,
Harvard-Smithsonian Center for Astrophysics, Cambridge MA 02135}
\email{j.ruostekoski@herts.ac.uk}

\date{\today}

\begin{abstract}
We perform a systematic study of the accuracy of split-step Fourier
transform methods for the time dependent Gross-Pitaevskii equation
using symbolic calculation. Provided the most recent
approximation for the wave function is always used in the nonlinear
atom-atom interaction potential energy, every split-step algorithm
we have tried has the same-order time stepping error for the
Gross-Pitaevskii equation and the Schr\"odinger equation.
\end{abstract}

\pacs{02.70.Wz,02.70.Hm,03.75.Kk} \maketitle

It is now firmly established that the time-dependent and independent
versions and multi-component generalizations of the Gross-Pitaevskii
equation (GPE) give a useful basic picture of most of the phenomena
taking place in a dilute atomic Bose-Einstein
condensate~\cite{DAL99,LEG01}. At this writing there are hundreds of
journal publications involving direct numerical solutions of the
GPE. The same equation, known as the nonlinear Schr\"odinger
equation, is a central topic in nonlinear optics~\cite{AGR95}.

Curiously, many research groups have their own unique numerical
methods, the methods are often ad-hoc, and the convergence
properties are not always stated (known?) explicitly. Even if
we restrict the discussion to the usual time dependent GPE,
there is enough variety to thwart our attempts at a classification;
but the Crank-Nicholson scheme and various split-step methods are
common themes~\cite{PRE00}. Our focus is on the analogs of the
classic split-step Fast Fourier Transform method for the
Schr\"odinger equation~\cite{FEI82}. The idea is to split the GPE
equation into two parts, one of which only refers to the momentum
operator and the other only to the position operator. The wave
function is then evolved alternatingly in momentum space and in real
space.

Nonetheless, the original split-operator method~\cite{FEI82} is
formally based on the algebra of linear operators, and does not
directly go over to the nonlinear GPE. Here we study the accuracy of
split-operator methods for the GPE using power series
expansions in the time step. While straightforward in principle,
with increasing order such expansions rapidly becomes unwieldy,
and are only manageable using symbolic calculation. Our surprise
discovery is that there is a simple and computationally inexpensive
way to include the nonlinearity so that the split-operator method
works for the GPE, and all the favorable properties of the original
algorithm~\cite{FEI82} that have made it a huge success are
preserved.

We first recap the case of the time-dependent Schr\"odinger
equation, for the time being in one spatial dimension. For brevity
of the notation we use units such that the mass of the particle and
the constant $\hbar$ both equal unity. The problem is to integrate
\begin{align}
&i {\partial\over\partial t}\,\psi(x,t) =[T +
V]\psi(x,t)\,;\\
&T \equiv-{1\over 2}\,{\partial^2\over\partial
x^2},\quad V\equiv V(x)\,.
\end{align}
A time step
from
$t$ to
$t+h$ is carried out formally as
\begin{equation}
\psi(x,t+h)=e^{-ih(T+V)}\psi(x,t)\,.
\label{SCHRSTEP}
\end{equation}
In position representation $e^{-ihV}$ means multiplication by the
function $e^{-ihV(x)}$. Similarly, in momentum representation, after
the Fourier transform ${\cal F}[\psi(x,t)]\rightarrow
\tilde\psi(p,t)$, the kinetic-energy exponential multiplies the wave
function $\tilde\psi(p,t)$ by $e^{-ihp^2/2}$. However, the operators
$T$ and $V$ do not commute, so that the inequality $e^{-ih(T+V)}\ne
e^{-ihV} e^{-ihT}$ holds and the exponential of the sum of the two
operators may be difficult to calculate.

Split-step methods attempt to get past the obstacle of noncommuting
operators by approximating
\begin{equation}
e^{\lambda(A+B)}\simeq e^{\lambda \beta_n B}e^{\lambda \alpha_n
A}\ldots e^{\lambda \beta_1 B}e^{\lambda \alpha_1 A}.
\label{SPLIT}
\end{equation}
In the present discussion it does not matter what the linear operators
are, so we refer to generic $A$ and $B$.
In practice the split~(\ref{SPLIT}) is not useful unless
there is an easy way to calculate each operator exponential on the
right-hand side, but this too is immaterial in our
formal development. Finally, we need not be dealing with time
stepping, thus we write the scalar parameter as $\lambda$. In general,
though, we regard $\lambda$ as small in absolute value. The idea of
split-operator methods is to pick the coefficients
$\alpha_i$ and $\beta_i$ so that the right-hand side of
the split~(\ref{SPLIT}) approximates the left-hand side to as high an
order in $\lambda$ as possible.
\begin{table}
\caption{\label{TAB:COEFFS} Nonzero coefficients for minimal
split-step methods with real coefficients for the orders of error
${\cal O}(\lambda^3)$,${\cal O}(\lambda^4)$, and ${\cal
O}(\lambda^5)$. The temporary notations are
$\Gamma=\sqrt{16-48\gamma+45\gamma^2-12\gamma^3\over9-12\gamma}$,
$\xi={1\over2-\sqrt[3]{2}}$, and $\gamma$ is a free real parameter
restricted so that $\Gamma$ remains real.}
\begin{ruledtabular}
\begin{tabular}{llll}
       & ${\cal O}(\lambda^3)$&${\cal
O}(\lambda^4)$&${\cal O}(\lambda^5)$
\\
    $\alpha_1$ & \half
&$1-{3\gamma-4/3\mp\Gamma\over2\gamma(\gamma\mp\Gamma)}$&\half$\xi$\\
    $\beta_1$ & 1 &$\gamma\mp\Gamma\over2$& $\xi$\\
    $\alpha_2$ & \half
&${3-4\gamma\over2(2-3\gamma)}$&${1-\sqrt[3]{2}\over2}\,\xi$\\
    $\beta_2$  &       &$\gamma\pm\Gamma\over2$&$-\sqrt[3]{2}\,\xi$\\
    $\alpha_3$ &
&${4/3-\gamma\pm\Gamma\over2\gamma(\gamma\pm\Gamma)}$&${1-\sqrt[3]{2}\over2}\,\xi$\\
    $\beta_3$ &        &1-$\gamma$&$\xi$\\
    $\beta_4$ &                          & &\half$\xi$\\
\end{tabular}
\end{ruledtabular}
\end{table}

We seek split-operator methods using symbolic manipulation on
{\it Mathematica}~\cite{WOL04}. We first expand the exponentials,
taking care not to inadvertently swap the noncommuting operators $A$
and
$B$. Next, in order to compare the left and right sides of
Eq.~(\ref{SPLIT}), it is expedient to put the operators into a
standard order, e.g., all operators $B$ to the left and all
operators $A$ to the right. This introduces commutators of the
operators, as in $AB = BA + [A,B]$. The difference between the left
and right sides of Eq.~(\ref{SPLIT}) is then arranged into a power
series of the parameter $\lambda$, where the coefficients contain
commutators of the operators and the constants $\alpha_i$,
$\beta_i$. The requirement that the difference cancels order by
order in $\lambda$ gives multivariate polynomial equations in
$\alpha_i$ and $\beta_i$, which {\it Mathematica} appears to solve
easily and completely.

The simplest nontrivial split-operator method found in this way is the
original split-operator algorithm~\cite{FEI82} with three
exponentials. The general operator-algebra argument does not
distinguish between the operators $A$ and $B$, and the choices for
the nonzero coefficients
$\alpha_1=\alpha_2=1/2$, $\beta_1=1$ and $\beta_1=\beta_2=1/2$,
$\alpha_2=1$ will both do. We have, for instance, a split-operator
representation
\begin{align}
&e^{\displaystyle{-ih\big[-{1\over2}{\partial^2\over\partial
x^2}+V(x)\big] }}\nonumber\\
=&e^{\displaystyle{ih\over4}{{\partial^2\over\partial
x^2}}}\,
e^{\displaystyle-{ih V(x)}}\,
e^{\displaystyle    {ih\over4}{{\partial^2\over\partial
x^2}}}+{\cal O}(h^3)\,.
\end{align}
Time stepping as in Eq.~(\ref{SCHRSTEP}) is readily implemented
using the Fast Fourier Transformation. The ensuing algorithm
automatically preserves the norm of the wave function, which is an
issue when standard differential equation solvers are applied to the
time-dependent Schr\"odinger equation. The implementation of the
exponential of the kinetic energy is a high-order spectral method,
and the exponential with the potential energy is done exactly in
principle.

A five-exponential split exists with an error ${\cal O}(\lambda^4)$,
but the coefficients $\alpha_i$ and $\beta_i$ are not real and the
algorithm is not absolutely stable. There is a whole family of
six-exponential splits with a continuous-valued free parameter that
all have an error ${\cal O}(\lambda^4)$, and the coefficients are
real for a range of the values of the free parameter. It takes a
seven-exponential split to gain another reduction of the error to
${\cal O}(\lambda^5)$, and higher-order methods are also found
without difficulty. However, for the present purpose we stop at
${\cal O}(\lambda^5)$, and list in Table~\ref{TAB:COEFFS} the
coefficients $\alpha_i$, $\beta_i$ for all minimal (as few
exponential factors as possible for a given order) split-operator
methods with real coefficients up to the order ${\cal
O}(\lambda^5)$. The first publication known to us that gives these
coefficients is Ref.~\cite{BAN91}.

On the other hand, the Gross-Pitaevskii equation
(GPE), or the nonlinear Schr\"odinger equation, reads in suitable
units
\begin{equation}
i{\partial\over\partial t}\psi = \left[-{1\over
2}{\partial^2\over\partial x^2}+V(x) + g|\psi|^2\right]\psi\,,
\end{equation}
where $g$ (usually $>0$) encapsulates the strength of the
(usually repulsive) atom-atom interactions. The GPE is nonlinear and
as such it does not fit the operator algebra framework we have used
to derive split-operator methods, but the square of the wave function
formally behaves like a potential energy. One is tempted to add it
to the potential-energy term in the time stepper, as in
\begin{eqnarray}
\lefteqn{\psi(x,t+h)}\nonumber\\
&\simeq&e^{\displaystyle{ih\over4}{{\partial^2\over\partial x^2}}}\,
e^{\displaystyle-{ih [V(x)\!+\!g|\psi(x,t)|^2]}}\, e^{\displaystyle
{ih\over4}{{\partial^2\over\partial x^2}}}\,\psi(x,t).\nonumber\\
\label{TENTSTEP}
\end{eqnarray}
\commentout{
\begin{align}
&\psi(x,t+h)\simeq\nonumber\\
& e^{\displaystyle{ih\over4}{{\partial^2\over\partial x^2}}}
e^{\displaystyle-{ih [V(x)\!+\!g|\psi(x,t)|^2]}}\, e^{\displaystyle
{ih\over4}{{\partial^2\over\partial x^2}}}\psi(x,t).
\label{TENTSTEP}
\end{align}}
Nonetheless,  as the wave function evolves in time, it is not
clear what to insert for the wave function in the exponential. In
fact, doing the step as in Eq.~(\ref{TENTSTEP}) drops an order in
accuracy, and the error turns out to be ${\cal O}(h^2)$.

In an attempt to find a better method we split the split-operator
step explicitly,
\begin{eqnarray}
\psi_0 &=& \psi(x,t);\nonumber\\
\psi_1 &=& e^{\displaystyle {ih\over4}{{\partial^2\over\partial
x^2}}}\psi_0;\nonumber\\
\psi_2 &=& e^{\displaystyle-{ih
[V(x)+g|c_0\psi_0+c_1\psi_1|^2]}}\psi_1;\nonumber\\
\psi(x,t+h) &=& e^{\displaystyle    {ih\over4}{{\partial^2\over\partial
x^2}}}\psi_2\,.
\label{SPLITEXP}
\end{eqnarray}
At the stage when the square of the wave function is
needed, there are already two versions of the wave function available.
The hope is to pick a linear combination of the two with the so far
unknown coefficients $c_0$ and $c_1$ to regain the error ${\cal
O}(h^3)$.

In order to produce the ``exact'' result for comparisons we write
\begin{equation}
\psi(x,t+h) = \psi(x,t) + {h\over1!}\,{\partial\over\partial t}
\psi(x,t)+
{h^2\over2!}\,{\partial^2\over\partial t^2}\psi(x,t)+\ldots.
\end{equation}
The $n$th time derivative of the wave function is obtained
inductively, by taking the ($n-1$)th time derivative of the GPE and
using the already existing expressions for the derivatives up to the
order $n-1$ to eliminate all time derivatives from the right-hand
side.  This procedure is carried out using {\it Mathematica},
treating real and imaginary parts of the wave function separately. A
mechanical implementation is well advised, as the terms multiply
rapidly with the order $n$; after complete expansion of the
derivatives, the real part of the fourth time derivative already has
168 linearly independent terms. The split-operator algorithm to be
tested is implemented by expanding the exponentials into power
series of the operators, and acting the series-form operators on the
initial wave function in the prescribed sequence. Such expansions
also become extremely tedious when the order in $h$ increases, and
they are done using {\it Mathematica}. The result is that the
method~(\ref{SPLITEXP}) gives an error ${\cal O}(h^3)$ if and only
if the coefficients $c_0$ and $c_1$ satisfy $c_0=0$, $|c_1|=1$; in
other words, if the most recent available wave function is used in
$g|\psi|^2$.

We have done a similar analysis for all split-operator methods with
the coefficients listed in Table I, starting  with both the position
step and the momentum step, and the result was the same every time:
If the most recently available version of the wave function is used
whenever $g|\psi|^2$ is needed alongside the potential energy, the
split-operator method for the GPE has the same order of accuracy in
the time step as the corresponding split-operator method for the
Schr\"odinger equation. Whether using the most recent update of the
wave function is also a necessary condition for the same order of
time stepping error in the linear and in the nonlinear problem
depends on the split-operator method on hand. For instance, if one
starts the three-split method with a position step and writes
\begin{eqnarray}
\psi_0 &=& \psi(x,t);\nonumber\\
\psi_1 &=& e^{\displaystyle-{1\over2}{ih
[V(x)+g|\psi_0|^2]}}\psi_0;\nonumber\\
\psi_2 &=& e^{\displaystyle {ih\over2}{{\partial^2\over\partial
x^2}}}\psi_1;\nonumber\\
\psi(x,t+h) &=& e^{\displaystyle-{1\over2}{ih
[V(x)+g|c_0\psi_0+c_1\psi_1+c_2\psi_2|^2]}}\psi_1\,,\nonumber\\
\label{SPLITALT}
\end{eqnarray}
all choices with $c_2=\pm1$ and $c_1=-c_0$ give an error  ${\cal
O}(h^3)$.

In the seven-step ${\cal O}(h^5)$ method on the order of 20000 terms
must identically cancel to validate our result, so clearly it is
not an accident. We conjecture that our observation holds for every
minimal split-step method, for arbitrary high orders. Also, so far
our examples have been in one spatial dimension. The split-operator
algorithms for the Schr\"odinger equation work
for arbitrary operators $A$ and $B$, and go over unchanged
to any number of spatial dimensions. Replacing the second
derivative in position with a multidimensional Laplacian should not
change anything in the underlying structure of the
split-operator method or the exact result in the GPE case either.
Our best guess is that the split-operator algorithms for the GPE work
the same way in more than one spatial dimension. We have verified
this explicitly for the ${\cal O}(h^3)$  three-exponential splits in
two and three spatial dimensions.

Our observation probably originates from some algebraic structure in
the GPE, but at this time we cannot tell what structure and how.
We see potential here for path-breaking insights into
norm-preserving nonlinear differential equations, and maybe
finite-difference equations as well. Moreover, computationally
demanding physics applications of the GPE arise with sets
of coupled GPEs encountered in multi-component
condensates\commentout{
\cite{RUO03}} and in atom-molecule coupling. We wonder if the
last-update rule could apply quite generally to norm-preserving
algorithms for such problems. One could
investigate this in any given special case using {\it Mathematica},
but understanding the underlying mathematics could lead to more
penetrating answers. Integrating the GPE and related
equations in imaginary time, a frequently employed method to find
the ground state for classical nonlinear fields, invites additional
interesting and possibly important questions of the same ilk.

In conclusion, by using the most recent existing update for the wave
function whenever the square of the wave function is needed
alongside the potential energy, in all of the cases we have studied
the split-step Fourier transform method goes over from the
Schr\"odinger equation to the Gross-Pitaevskii equation with its
favorable properties intact. We hope that our observation is useful
in future studies based on the Gross-Pitaevskii equation and its
variations and generalizations.

This work is supported in part by NSF (PHY-0354599), NASA
(NAG3-2880), EPSRC, and NSF through a grant for the Institute for
Theoretical Atomic, Molecular and Optical Physics at Harvard
University and Smithsonian Astrophysical Observatory.

\end{document}